

\magnification \magstep1
\hoffset 1.5truecm
\hsize 16truecm \vsize 23.5truecm
\baselineskip 20pt \parskip5pt
\raggedbottom

\def\section#1{\bigskip\penalty-200\centerline{\bf#1}
 \smallskip}

\def\Deg{{\rm Deg}}
\def\gh{{\rm gh}}
\def\Tr{{\rm Tr}}
\def\half{{1\over2}}


\centerline{\bf QUANTUM THEORY OF NON-ABELIAN DIFFERENTIAL
FORMS}
\centerline{\bf AND LINK POLYNOMIALS}

\centerline{BOGUS\L AW BRODA%
            \footnote{$^\star$}{Humboldt fellow.}}

{
\it
\centerline{Arnold Sommerfeld Institute for Mathematical
Physics}
\centerline{Technical University of Clausthal,
Leibnizstra\ss e 10}
\centerline{D-W--3392 Clausthal-Zellerfeld, Federal
Republic of
Germany}
\centerline{\rm and}
\centerline{Department of Theoretical Physics, University
of \L\'od\'z}
\centerline{Pomorska 149/153, PL--90-236 \L\'od\'z, Poland%
            \footnote{$^\dagger$}{\rm Permanent address.}}
}

\vfill

A topological quantum field theory of non-abelian
differential forms is investigated from the point of view
of its possible applications to description of polynomial
invariants of higher-dimensional two-component links. A
path-integral representation of the partition function of
the theory, which is a highly on-shell reducible system, is
obtained in the framework of the antibracket-antifield
formalism of Batalin and Vilkovisky.  The quasi-monodromy
matrix, giving rise to corresponding skein relations, is
formally derived in a manifestly covariant non-perturbative
manner.

\vfill

\centerline{MARCH 1993}

\vfill\eject

\section{Introduction}
Metric-independent gauge theories with a non-trivial
classical action, so-called topological field theories of
the Schwarz type (see Ref.~1, for a comprehensive review of
all topological theories), play an important role in
``physics'' approach to polynomial invariants of knots and
links. The first paper$^2$ establishing this direction of
research deals with an application of Chern-Simons theory
to the Jones (or more generally to the Homfly) polynomial.
The second important example of a theory of this type is
provided by gauge theory of (non-abelian) differential
forms, so-called BF-theory.$^{3-8}$ Although the pure
BF-theory is now a well-established system, it seems that
the problem of physical observables and/or corresponding
possible topological invariants has not yet been fully
satisfactory solved in its context. For example, in Refs.~4
and~5 an additional restricting flatness condition is
imposed on the curvature, whereas Ref.~6 exclusively deals
with abelian observables, yielding gaussian linking
numbers. Ref.~7 only sketches an idea, and a particular
four-dimensional version of Ref.~8 does not explicitly
relate the observables introduced to possible topological
invariants. In this work we formally derive, in the
framework of BF-theory, skein relations for some
higher-dimensional two-component links.

Roughly speaking, there are two kinds of difficulties
related to BF-theory in an arbitrary dimension $d$.  The
first rather conceptual difficulty concerns the problem of
``physical observables'' measuring linking phenomena in
higher dimensions. In the standard (three-dimensional)
Chern-Simons case, one traditionally introduces the Wilson
loops, but obviously this method does not work in higher
dimensions.  To encode topologically interesting
information in the framework of higher-dimensional theory
one should introduce topological ``matter'' multiplets
living on the submanifolds corresponding to the
higher-dimensional knots/links under consideration. Since
the purpose of our paper is to derive skein relations for a
two-component link (the link consisting of the two
components, $\cal K$ and $\cal C$, of dimension $d-2$ and
$1$, respectively) we will supplement the standard
BF-action with two sets of topological matter
multiplets.$^7$

The second more technical difficulty concerns the problem
of covariant quantiza\-tion.$^9$ BF-theory, as a highly
on-shell reducible system, requires a treatment in the
framework of the antibracket-antifield formalism of Batalin
and Vilkovisky.$^{10}$ Solutions of this problem, at least
in the case of the pure BF-theory, are presented in Ref.~11
(see also Ref.~12).

In the first ``classical'' part of the work, we will define
a total classical action of the full theory, i.~e.
BF-system plus the topological matter part, derive
classical equations of motion and find classical symmetries
of the action. The second ``quantum'' part is devoted to
the BRST quantization of the system in the framework of the
formalism of Batalin and Vilkovisky. We will present an
explicit and elegant form of the solution of the master
equation, as well as a compact form of the BRST $s$
operator, playing here an auxiliary role. A covariant
path-integral representation of the partition function
appears as a straightforward consequence of the formalism
used. In the third ``topological'' part, appealing to
reader's imagination and using the Stokes theorem, we
derive (in a non-perturbative way) the quasi-monodromy
matrix giving rise to skein relations corresponding to
an arbitrary pair of irreducible representations of an
arbitrary compact semisimple Lie group $G$. Appendix
contains some useful formulas valid also in a more general
case.

\section{1. Classical action}
In $d$ dimensions ($d\geq2$), the classical action of gauge
theory of non-abelian differential forms (BF-theory) is
defined as$^6$
$$
S_{BF}^{\rm cl}={1\over\lambda}\int_{{\cal
S}}\Tr(BF),
\eqno(1.1)
$$
where the coupling constant $\lambda$ can assume an
arbitrary non-zero complex value, ${\cal S}$ is a
$d$-dimensional sphere (formally, one could also try to
consider a more general manifold), $B$ is a Lie-algebra
valued ($d-2$)-form, $B=T^aB^a$, $F$ is the curvature
two-form, and the normalization of the generators $T^a$ of
the compact semisimple Lie group $G$ is
$\Tr(T^aT^b)=\half\delta^{ab}$.  For differential forms all
products are exterior ones. It is interesting to note that
contrary to Chern-Simons theory the coupling constant
$\lambda$ is not constrained to integer values.
Accordingly, the corresponding parameter in skein relations
is not constrained to integer values either.

The classical action of matter part of gauge theory of
non-abelian differential forms consists of the two pieces
($d\geq3$)$^7$:

\noindent
(1)
$$
S_\Omega^{\rm cl}=\half\int_{{\cal K}}
(\bar\Theta d_A\Omega+ d_A\bar\Omega\Theta+\bar\Theta
B\Theta),
\eqno(1.2{\rm a})
$$
where $\Theta$ and $\bar\Theta$ are zero-forms, $\Omega$
and $\bar\Omega$ are ($d-3$)-forms (all the forms are in an
irreducible representation $R_1(G)$ with the generators
$t_1^a$), $ d_A$ is the exterior
covariant derivative,
$ d_A\Omega\equiv d\Omega+A\Omega$,
$ d_A\bar\Omega\equiv d\bar\Omega-A^T\bar\Omega$,
$A\equiv t_1^aA^a$,
and ${\cal K}$
is a ($d-2$)-dimensional closed submanifold imbedded in
${\cal S}$, a ($d-2$)-knot (the first component of the link
$\cal L$);

\noindent
(2)
$$
S_\eta^{\rm cl}=\half\int_{\cal C}\bar\eta d_A\eta,
\eqno(1.2{\rm b})
$$
where $\eta$ and $\bar\eta$ are zero-forms in an
irreducible representation
$R_2(G)$, and $\cal C$ is
a one-dimensional loop imbedded in $\cal S$, a standard
knot (the second component of the link~$\cal L$).

Then the classical action of the whole theory is given as
the sum
$$
S^{\rm cl}=S_{BF}^{\rm cl}+S_\Omega^{\rm cl}+S_\eta^{\rm
cl}.
\eqno(1.3)
$$
Hence the corresponding classical equations of motion are
of the form
$$
d_AB+\lambda(\bar\Omega
t_1\Theta-\bar\Theta t_1\Omega)\delta({\cal K})
-\lambda\bar\eta t_2\eta\delta({\cal C})=0,
$$
$$
F+\lambda\bar\Theta t_1\Theta\delta({\cal K})=0,
$$
$$
 d_A\Omega+B\Theta=0,\qquad d_A\bar\Omega+\bar\Theta B=0,
$$
$$
 d_A\Theta=0,\qquad d_A\bar\Theta=0,
$$
$$
 d_A\eta=0,\qquad d_A\bar\eta=0,
\eqno(1.4)
$$
where $\delta({\cal K})$ and $\delta(\cal C)$ is a
Dirac-delta two-form and a ($d-1$)-form, respectively (see
Appendix (1)).
A subset of all solutions of the classical equations of
motion, important for further symmetry analysis, is given
by
$$
 d_AB=F= d_A\Omega= d_A\bar\Omega
=\Theta=\bar\Theta=\eta=\bar\eta=0.
\eqno(1.5)
$$

In general case, i.~e.\ $d\geq4$, the action (1.3) enjoys
four kinds of local gauge symmetries:

\noindent
(1) Ordinary gauge symmetry ($d\geq2$)
$$
\delta_1A=-{1\over2\lambda} d_A\sigma_1
\equiv-{1\over2\lambda}( d\sigma_1+[A,\sigma_1]),
$$
$$
\delta_1B={1\over2\lambda}[\sigma_1,B],
$$
$$
\delta_1\Omega={1\over2\lambda}\sigma_1\Omega,\qquad
\delta_1\bar\Omega=-{1\over2\lambda}\bar\Omega\sigma_1,
$$
$$
\delta_1\Theta={1\over2\lambda}\sigma_1\Theta,\qquad
\delta_1\bar\Theta=-{1\over2\lambda}\bar\Theta\sigma_1,
$$
$$
\delta_1\eta={1\over2\lambda}\sigma_1\eta,\qquad
\delta_1\bar\eta=-{1\over2\lambda}\bar\eta\sigma_1,
\eqno(1.6)
$$
where $\sigma_1$ is a zero-form in $R_{\rm Adj}(G)$.

\noindent
(2) $B$-symmetry ($B$ stands for $B$-field or Bianchi)
$$
\delta_2A=\delta_2\Theta=\delta_2\bar\Theta
=\delta_2\eta=\delta_2\bar\eta=0,
\eqno(1.7{\rm a})
$$
$$
\delta_2B={1\over2\lambda} d_A\sigma_2,
\eqno(1.7{\rm b})
$$
$$
\delta_2\Omega=-{1\over2\lambda}\sigma_2\Theta,\qquad
\delta_2\bar\Omega=-{1\over2\lambda}\bar\Theta\sigma_2,
\eqno(1.7{\rm c})
$$
where $\sigma_2$ is a ($d-3$)-form in $R_{\rm Adj}(G)$.
This symmetry emerges for $d\geq3$, and is reducible for
$d\geq4$. The reducibility follows from the fact that we
can perform an additional transformation
$$
\delta_2^\prime\sigma_2={1\over2\lambda} d_A\sigma_2^\prime,
\eqno(1.8)
$$
which is an on-shell symmetry transformation of (1.7b).
Namely
$$
\delta_2^\prime\delta_2B
={1\over4\lambda^2} d_A^2\sigma_2^\prime
={1\over4\lambda^2}[F,\sigma_2^\prime]=0,
\eqno(1.9)
$$
where the solution (1.5) of the equations of motion has
been used. For $d\geq5$, we can repeat this procedure
performing further transformations
$$
\delta_2^{\prime\prime}\sigma_2^\prime
={1\over2\lambda} d_A\sigma_2^{\prime\prime},
\eqno(1.10)
$$
which is an on-shell symmetry transformation of (1.8),
and so on. We conclude, that according to the
Batalin-Vilkovisky quantization scheme,$^{10}$ we deal with
($d-3$)-stage on-shell reducible gauge symmetry.

\noindent
(3) ``Matter'' gauge symmetry of $\Omega$
$$
\delta_3A=\delta_3\bar\Omega=\delta_3\Theta=\delta_3\bar\Theta
=\delta_3\eta=\delta_3\bar\eta=0,
\eqno(1.11a)
$$
$$
\delta_3B=\half\lambda\bar\Theta
t_1\sigma_3\delta({\cal K}),
\eqno(1.11b)
$$
$$
\delta_3\Omega=-\half d_A\sigma_3,
\eqno(1.11c)
$$
where $\sigma_3$ is a ($d-4$)-form in $R_1(G)$. This
symmetry emerges for $d\geq4$, and it appears to be
reducible for $d\geq5$. The obvious on-shell symmetry
transformation of (1.11c) is
$$
\delta_3^\prime\sigma_3=-\half d_A\sigma_3^\prime.
\eqno(1.12)
$$
Accordingly, the symmetry is ($d-4$)-stage on-shell
reducible.

\noindent
(4) ``Matter'' gauge symmetry of $\bar\Omega$
$$
\delta_4A=\delta_4\Omega=\delta_4\Theta=\delta_4\bar\Theta
=\delta_4\eta=\delta_4\bar\eta=0,
$$
$$
\delta_4B=\half\lambda\delta({\cal K})
\bar\sigma_4t\Theta,
$$
$$
\delta_4\bar\Omega=-\half d_A\bar\sigma_4,
\eqno(1.13)
$$
where $\bar\sigma_4$ is a ($d-4$)-form in $R_1(G)$. This
symmetry is, in general, also ($d-4$)-stage on-shell
reducible.

We have chosen a bit strange normalization of our gauge
transformations because we want to achieve exact agreement
with the non-degeneracy condition (A2.3).  One should also
note that the $B$-symmetry would be broken out if
$\partial\cal K\neq\emptyset$, and the ``matter''
symmetries would be broken out by the last term in (1.2a)
if the submanifold $\cal K$ had self-intersections. But
since we would like to interpret $\cal K$ as a knot we
should assume that $\cal K$ is closed, and that it has no
self-intersections.

\section{2.~Quantum action}
Covariant quantizing on-shell reducible gauge systems can
be approached by means of the Batalin-Vilkovisky
antifield-antibracket procedure.$^{10}$  The final result
of such a procedure is a covariant path-integral
representation of the partition function $Z$. The problem
is essentially solved if one succeeds in finding a proper
non-degenerate solution $S$ (an extended classical action,
or a classical part of the quantum action $W$) of the
master equation
$$
(S,S)=0,
\eqno(2.1)
$$
where the antibracket is defined in Appendix (2).

Instead of trying to solve the master equation
perturbatively we simply postulate that the solution is of
the form analogous to the form of the classical action
$S^{\rm cl}$ (strictly speaking, a minimal part of $S$),
and then it is given by
$$
S=S_{BF}+S_\Omega+S_\eta,
\eqno(2.2)
$$
with
$$
S_{BF}={1\over\lambda}\int_{{\cal S}}\Tr(bf),
\eqno(2.3{\rm a})
$$
$$
S_\Omega=\half\int_{\cal
K}(\bar\theta d_a\omega+ d_a\bar\omega\theta
+\bar\theta b\theta)
=\half\int_{\cal K}\left(\bar\theta d_A\omega
+ d_A\bar\omega\theta+\bar\Theta B\Theta\right),
\eqno(2.3{\rm b})
$$
where
$$
 d_a\omega\equiv d\omega+a\omega,
$$
$$
 d_a\bar\omega\equiv d\bar\omega-a^T\bar\omega,
$$
$$
f\equiv d_a^2\equiv d a+a^2,
\eqno(2.4)
$$
and
$$
S_\eta=\half\int_{\cal C}\bar\eta d_a\eta
=\half\int_{\cal C}\bar\eta d_A\eta=S_\eta^{\rm cl}.
\eqno(2.3{\rm c})
$$
The fields entering (2.3) are five-graded (four-graded with
respect to the four ghost numbers corresponding to the four
gauge symmetries) non-homogeneous forms in respective
representations of $G$. They constitute a minimal sector of
the theory, and they assume the following explicit form:
$$
a\equiv\sum_{g=0}^1A_{1-g}^{g\ 0\ 0\
0} +\sum_{g=0}^{d-2}B_{g+2}^{\ast\ 0\ -1-g\ 0\ 0},
$$
$$
b\equiv\sum_{g=0}^1A_{d-1+g}^{\ast\ -1-g\ 0\ 0\ 0}
-\sum_{g=0}^{d-2}B_{d-2-g}^{0\ g\ 0\ 0},
$$
$$
\omega\equiv\sum_{g=0}^{d-3}\Omega_{d-3-g}^{0\ 0\ g\ 0},
\qquad
\bar\omega\equiv\sum_{g=0}^{d-3}\bar\Omega_{d-3-g}^{0\ 0\
0\ g},
$$
$$
\theta\equiv\Theta+\sum_{g=1}^{d-2}\bar\Omega_g^{\ast\ 0\ 0\ 0\
-g},
\qquad
\bar\theta\equiv\bar\Theta+\sum_{g=1}^{d-2}\Omega_g^{\ast\ 0\ 0\
-g\ 0},
\eqno(2.5)
$$
where the following identifications have been assumed for
the classical fields
$\phi^{\rm cl}=\{A,B,\Omega,\bar\Omega,\Theta,\bar\Theta\},$
$$
A\equiv A_{1}^{0\ 0\ 0\ 0},
\qquad
B\equiv B_{d-2}^{0\ 0\ 0\ 0},
$$
$$
\Omega\equiv\Omega_{d-3}^{0\ 0\ 0\ 0},
\qquad
\bar\Omega\equiv\bar\Omega_{d-3}^{0\ 0\ 0\ 0},
$$
$$
\Theta\equiv \Theta_0^{0\ 0\ 0\ 0},
\qquad
\bar\Theta\equiv\bar\Theta_0^{0\ 0\ 0\ 0},
\eqno(2.6)
$$
and for the minimal sector $\phi^{\rm min}$
$$
A^{\rm min}
=\left\{A_{1}^{0\ 0\ 0\ 0},A_{0}^{1\ 0\ 0\ 0}\right\},
$$
$$
B^{\rm min}
=\left\{B_{d-2}^{0\ 0\ 0\ 0},B_{d-3}^{0\ 1\ 0\ 0},
\dots,B_{0}^{0\ d-2\ 0\ 0}\right\},
$$
$$
\Omega^{\rm min}=\left\{\Omega_{d-3}^{0\ 0\ 0\ 0},
\Omega_{d-4}^{0\ 0\ 1\ 0},\dots,
\Omega_{0}^{0\ 0\ d-3\ 0}\right\},
$$
$$
\bar\Omega^{\rm min}=\left\{\bar\Omega_{d-3}^{0\ 0\ 0\ 0},
\bar\Omega_{d-4}^{0\ 0\ 0\ 1},
\dots,\bar\Omega_{0}^{0\ 0\ 0\ d-3}\right\}.
\eqno(2.7)
$$
The total degrees ``Deg'' (A3.4) of our forms are
$$
\Deg a=\deg A=1,
$$
$$
\Deg b=\deg B=d-2,
$$
$$
\Deg\omega=\Deg\bar\omega=\deg\Omega=\deg\bar\Omega=d-3,
$$
$$
\Deg\theta=\Deg\bar\theta=\deg\Theta=\deg\bar\Theta=0.
\eqno(2.8)
$$
It is implicitly assumed that only the integrands with all
ghost numbers equal to zero, and with form degrees equal to
the dimensions of respective manifolds survive in (2.3).

Now, we can introduce a BRST operator $s$. To preserve
self-consistency, the action of $s$ on the fields should be
defined according to (A3.2). In a compact notation,
$$
sa={1\over2\lambda}f+\half
\bar\theta t_1\theta\delta({\cal K}),
$$
$$
sb={1\over2\lambda} d_ab+\half(\bar\omega t_1\theta-\bar\theta
t_1\omega)\delta({\cal
K})-\half\bar\eta t_2\eta\delta({\cal C}),
$$
$$
s\omega=\half( d_a\omega+b\theta),
$$
$$
s\bar\omega=\half( d_a\bar\omega+\bar\theta b),
$$
$$
s\theta=(-)^{d+1}\half d_a\theta,
$$
$$
s\bar\theta=(-)^{d+1}\half d_a\bar\theta,
$$
$$
s\eta=s\bar\eta=0,
\eqno(2.9)
$$
where
$$
s\Theta=s\bar\Theta=0.
\eqno(2.9{\rm a})
$$
Performing a very straightforward calculation we can easily
check
that (see (A3.1))
$$
(S,S)\equiv sS=0,
\eqno(2.10)
$$
provided the obvious additional topological condition  $\cal
K\cap C=\emptyset$ is satisfied. To perform the
calculation one should make use of the Stokes theorem, the
generalized Bianchi identity,
$$
d_af\equiv0,
\eqno(2.11)
$$
as well as the formulas (A3.5).  It appears that our BRST
operator $s$ is automatically nilpotent, $s^2=0$ (see
(A3.3)).  Also, it can be easily checked that $S$ possesses
the correct classical limit in the sense of Batalin and
Vilkovisky
$$
S(\phi,\phi^\ast)\vert_{\phi^\ast=0}=S^{\rm cl}(\phi^{\rm cl}),
\eqno(2.12)
$$
where the collective symbol $\phi^\ast$ denotes all
antifields, and that it satisfies the condition of
non-degeneracy (A2.3)
$$
\left.{\delta_l\delta_rS\over
\delta A_{d-1}^\ast\delta A_0}\right\vert_{\phi^\ast=0}
=-{1\over2\lambda}\ast d_A\delta,
$$
$$
\left.{\delta_l\delta_rS\over
\delta B_{1+g}^\ast\delta B_{d-2-g}}\right\vert_{\phi^\ast=0}
={1\over2\lambda}\ast d_A\delta,
\qquad1\leq g\leq d-2,
$$
$$
\left.{\delta_l\delta_rS\over
\delta\Omega_g^\ast\delta\Omega_{d-3-g}}
\right\vert_{\phi^\ast=0}=-\half\ast d_A\delta,
\qquad1\leq g\leq d-3,
$$
$$
\left.{\delta_l\delta_rS\over
\delta\bar\Omega_g^\ast\delta\bar\Omega_{d-3-g}}
\right\vert_{\phi^\ast=0}=-\half\ast d_A\delta,
\qquad1\leq g\leq d-3,
\eqno(2.13)
$$
where the first two covariant derivatives act in the
adjoint representation $R_{\rm Adj}(G)$, the second two
ones in $R_1(G)$, and $\delta$ denotes the ordinary
Dirac-delta.

The extended classical action should be supplemented
with the auxiliary term $S^{\rm aux}$. The form of $S^{\rm
aux}$ is universally given for an arbitrary theory in Ref.~10,
and in our case
$$
S^{\rm aux}\left(\phi_{\rm aux}^{\ast},\Pi^{\phi}\right)
=\int_{{\cal S}}\Tr(A_{\rm aux}^{\ast}\Pi^{A}
+\sum_iB_{\rm aux}^{\ast\ i}\Pi_{i}^{B})
+\half\int_{\cal K}\sum_j(
\Omega_{\rm aux}^{\ast\ j}\Pi_{j}^{\Omega}
+\bar\Omega_{\rm aux}^{\ast\ j}\Pi_{j}^{\bar\Omega}),
\eqno(2.14)
$$
where
$\Pi^\phi=\{\Pi^A,\Pi_i^B,\Pi_j^\Omega,\Pi_j^{\bar\Omega}\}$
consists of a multiplet of Lagrange multipliers
(Nakani\-shi-Lautrup-Stueckelberg fields), and $\phi_{\rm
aux}^{\ast}=\{A_{\rm aux}^{\ast},B_{\rm aux}^{\ast\
i},\Omega_{\rm aux}^{\ast\ j},\bar\Omega_{\rm aux}^{\ast\
j}\}$ denotes a multiplet of antifields in the
auxiliary sector. Thus all the fields appearing in the full
action $S$, minimal and auxiliary, constitute the four
triangles of ghosts corresponding to the four gauge
symmetries. The explicit form of the form degrees and of
the ghost numbers uniquely follows from the
Batalin-Vilkovisky prescription and the duality condition.
{}From (A3.2), we directly obtain
$$
s\phi_I^{\rm aux}=\half\Pi_I^\phi,
$$
$$
s\phi_{\rm aux}^\ast=s\Pi=0,
\eqno(2.9b)
$$
where $I$ denotes a respective index.

One should stress that our extended classical
action satisfies not only the classical master equation
(2.1) but also the quantum one
$$
\half(S,S)= i\Delta_{\rm BV}S\equiv i{\delta_r\delta_lS
\over\delta\phi^I\delta\phi_I^\ast}=0.
\eqno(2.15)
$$
To check it we can observe that the RHS of (2.15) vanishes
identically by virtue of antisymmetry of the structure
constants. Strictly speaking, the RHS of (2.15) is not
well-defined mathematically, but it already vanishes for a
regularized version. Hence the full quantum action
$W=S+S^{\rm aux}$.

As a final step of the quantization procedure we should
define the gauge fermion $\Psi(\Phi)$ that satisfies some
conditions of non-degeneracy. In turn, the gauge fermion
defines the antifields,
$$
\phi_I^\ast=\ast{\delta_r\Psi\over\delta\phi^I}.
\eqno(2.16)
$$
Since we are not going to perform any perturbative
calculations we need not to fix a concrete form of $\Psi$.
In fact, our further analysis is independent of a
particular form of $\Psi$, and exclusively rests only on
its existence.

Thus the partition function of our theory can be written in
the following covariant path-integral representation
$$
Z=\int D\phi D\Pi\exp( i W)\vert_\Sigma,
\eqno(2.17)
$$
where $\phi=\left\{\phi^{\rm min},\phi^{\rm aux}\right\}$,
and the symbol $\Sigma$ indicates that we should
eliminate antifields using (2.16).

As the gauge fixing procedure (the gauge fermion $\Psi$)
unavoidably introduces the metric tensor, the question
arises as to the metric independence of $Z$.$^9$ Since the
metric tensor enters $Z$ only through $\Psi$ it follows
from the theorem of Batalin and Vilkovisky (on
gauge-independence) that $Z$ should also be metric
independent.

\section{3.~Invariant polynomials}
In this section, we will approach the problem of invariant
polynomials for higher-dimensional links consisting of the
two components, $\cal K$ and $\cal C$, of the dimension
$d-2$ and~1, respectively. From topological point of view,
our approach is purely formal, and presumably it concerns
only some subclass of genuinely higher-dimensional
invariants of links, where possibly some matrices
corresponding to the simplex equation rather than to the
Yang-Baxter one would appear. However, our approach is
fully motivated, as we consider non-trivial physical
observables, which describe some linking phenomena in $d$
dimensions in a non-trivial way.

The prototype of our two-component link is the pair
consisting of the closed manifold ${\cal K}$ and the loop
$\cal C$, which enter our theory through the topological
matter action $S_\Omega+S_\eta$.  As we would like to
derive the topological invariants in the form of invariant
polynomials we should derive corresponding skein relations.
To this end, we have to compare some number of copies
${\cal L}_k$ ($k=1,2,\dots,N$) of the link $\cal L$,
entering our skein relation, appropriately differing inside
$d$-dimensional balls ${\cal B}$ ($N\geq3$, and $N$ depends
on the pair of irreducible representations, $R_1(G)$ and
$R_2(G)$. An explanation of the word ``appropriately''
implicitly follows from our further construction. To
calculate the contributions to the functional integral
coming from ${\cal B}$'s that are different for the
different copies ${\cal L}_k$ we will use the Stokes
theorem. First of all, we postulate that each $\cal B$
contains two connected (but obviously disjoint) pieces of
the corresponding copy of $\cal L$, say $\cal K^\prime$ and
$\cal C^\prime$. For simplicity, suppose that only the
pieces $\cal C^\prime$ are different, whereas $\cal
K^\prime$ are identical for all ${\cal L}_k$.  In order to
facilitate comparison of the different situations one
should assume some standard position of $\cal C^\prime$ in
each $\cal B$. It is a two-dimensional surface $\cal D$
inside each $\cal B$ swept out by $\cal C^\prime$ that
differs $\cal C$ for different copies of~$\cal L$.
Analytically, the difference is expressed by the following
integral
$$
\Delta S=\half\int_{\cal C^\prime=\partial D}
\bar\eta d_A\eta
=\half\int_{\cal D}( d_A\bar\eta d_A\eta
+\bar\eta F\eta),
\eqno(3.1)
$$
where the Stokes theorem has been used. Now we suppose that
the intersection of $\cal D$ and $\cal K^\prime$ is exactly
in $k-1$ points for each ${\cal L}_k$. As the theory is
topological (metric-independent) we can deform the links
freely without any influence on the path integral, provided
we avoid intersections, which constitute a
kind of singularities. The contributions coming from the
intersections can be easily calculated, and they are the
only analytical trace of our construction. Since the
dimension of $\cal K^\prime$ is $d-2$, and the dimension of
$\cal D$ is $2$, the dimension of $\cal K^\prime\cap D$ is,
in general position, $0$. In fact, we can assume
$$
{\cal K^\prime\cap D}=\cases{
\emptyset,&for $k=1$,\cr
\bigcup_{l=1}^{k-1}{\cal P}_l,&for $2\leq k\leq N$,\cr}
\eqno(3.2)
$$
where ${\cal P}_l$ are intersection points. Now, we can
perform the following substitution in (3.1)
$$
F^a\longrightarrow-2 i\lambda\ast{\delta\over\delta B^a},
\eqno(3.3)
$$
provided the order of terms in (2.17) is such that the
functional derivative (3.3) can act on $S_{BF}$ yielding
the curvature $F$. We can observe that (3.3) is a
translation operator in a function space. Functionally
integrating by parts with respect to $B$ we obtain, as a
result of a translation in the last term of $S_\Omega$, the
quasi-monodromy ``operator''
$$
M=\exp\left[{\lambda\over2 i}
(\bar\theta t_1^a\theta)(\bar\eta t_2^a\eta)(x_l)\right],
\eqno(3.4)
$$
for each ${\cal P}_l$ with coordinates $x_l$ (summation
with respect to $a$).

Since there are other terms entering the path integral that
could possibly give a contribution to (3.4), one should
explain why this is not the case. In particular, only the
``potential'' term in (3.1) is expected to give a
contribution to the path integral (exactly in the form
(3.4)). To understand this fact we can think of a lattice
formulation of our theory, where $\cal D$ is a plaquette.
As usual, $F$ should live on plaquettes, whereas $A$ should
live on bonds. Since $\cal K$ pierces the plaquette $\cal
D$ rather than a bond there should be no contribution from
the ``kinetic term'', which just resides on the bond. One
can also observe that the field $B$ enters as well the
gauge fermion $\Psi$, and therefore $\Psi$ could be
affected by (3.3), but due to the theorem of Batalin and
Vilkovisky on the $\Psi$-independence of $Z$ this change of
$\Psi$ is inessential. Summing up, the whole contribution
coming from the intersection(s) is contained in the
monodromy operator ~(3.4).

Now, it is necessary to calculate matrix elements of $M$.
To this end, we should be provided with appropriate scalar
products. Obviously, these scalar products should be
already contained in the theory rather than given from
outside.  Tracing the standard method of the derivation of
path-integral representation of a partition function
from the operator formulation we can easily decipher the
form of the scalar products from the form of the
``kinetic'' terms. Namely, there are the following three
(normalized) non-trivial scalar products for the matter
fields:
$$
(f\cdot g)_{\bar\eta\eta}\equiv{1\over2\pi i}\int fg
e^{ i\bar\eta\eta} d\bar\eta d\eta,
\eqno(3.5{\rm a})
$$
$$
(f\cdot g)_{\bar\theta\omega}\equiv{1\over2\pi i}\int fg
e^{ i\bar\theta\omega} d\bar\theta d\omega,
\eqno(3.5{\rm b})
$$
$$
(f\cdot g)_{\bar\omega\theta}\equiv{1\over2\pi i}\int fg
e^{ i\bar\omega\theta} d\bar\omega d\theta.
\eqno(3.5{\rm c})
$$
The matrix elements of $M$ are given by the following
fourfold scalar product
$$
{\bf M}=(\bar\eta\prime\bar\omega\prime\cdot
M\cdot\omega\prime\eta\prime)
=\exp\left({\lambda\over2 i}t_1^a\otimes t_2^a\right).
\eqno(3.5)
$$
Interestingly, the algebraic form of {\bf M} is identical
to the standard, three-dimensional one. Since our approach
applies to three dimensions as well, we have obtained, as a
by-product of our analysis, the most general (where the
parameter $\lambda$ is unconstrained) form of the
quasi-monodromy matrix.

Having a particular compact semi-simple Lie group $G$ and a
pair of irreducible representations $R_1(G)$ and $R_2(G)$
we can automatically yield a corresponding skein relation,
following the recipe given in.$^{13}$ For example, for the
fundamental representations of SU($N$) we obtain the Homfly
polynomial, which specializes to the Jones polynomial after
putting $N=2$. For SO($N$) we obtain the Dubrovnik-Kauffman
polynomial, whereas non-fundamental representations provide
us, as a rule, with the Akutsu-Wadati polynomials.

{\noindent\bf
Acknowledgments}

The author is indebted to Prof. H.~D.~Doebner for his kind
hospitality in Clausthal. The work was supported by the
Alexander von Humboldt Foundation, the Polish grant
no KBN~202189101, and the grant of the Commission of the
European Communities no CIPA3510PL921596.

\section{Appendix}
\noindent
{\bf (1)}\quad
A Dirac-delta $n$-form $\delta(\cal N)$, $0<n<d$,
where $\cal N$ is a ($d-n$)-dimensional submanifold of a
$d$-dimensional manifold $\cal M$, is defined through the
relation
$$
\int_{\cal M}(\dots)\delta({\cal N})\equiv\int_{\cal N}(\dots),
\eqno({\rm A}1.1)
$$
where ``$\dots$'' denote some ($d-n$)-form, i.~e.
$\delta(\cal N)$ constraints integration on $\cal M$
to the submanifold $\cal N$. If we parametrize $\cal N$
with $\{x^{n+1},x^{n+2},\dots,x^d\}$ in a locally cartesian
coordinate system
$\{x^1,x^2,\dots,x^n,x^{n+1},\dots,x^d\}$, $\delta(\cal
N)$ will assume the following simple explicit local form
$$
\delta({\cal
N})=\delta(x^1)\delta(x^2)\dots\delta(x^n)\epsilon_{12\dots
n} d x^1\wedge d x^2\wedge\dots\wedge d x^n.
\eqno({\rm A}1.2)
$$
{}From (A1.2) it directly follows that
$$
\delta^2({\cal N})=0.
\eqno({\rm A}1.3)
$$
The square of the Dirac-delta $n$-form $\delta(\cal N)$ is
zero in a meaningful way because it vanishes already in a
regularized version due to antisymmetry of differential forms.
Presumably, a mathematically more rigorous description in
terms of de~Rham's currents would also be possible.$^{14}$
\bigskip
\noindent
{\bf(2)}\quad
In the framework of the antibracket-antifield formalism of
Batalin and Vilkovisky, the antibracket of arbitrary two
functions $X$ and $Y$ on the extended phase space of
variables $\{\phi^I,\phi_I^\ast\}$ is defined according to
Ref.~10 as
$$
(X,Y)\equiv{\delta_rX\over\delta\phi^I}
{\delta_lY\over\delta\phi_I^\ast}
-{\delta_rX\over\delta\phi_I^\ast}
{\delta_lY\over\delta\phi^I},
\eqno({\rm A}2.1)
$$
where, in the condensed notation, $I=(i,x)$ denotes discrete
as well as continuous indices, and $r$ ($l$) means right
(left) derivative. We assume that form degrees of antifields
are determined by duality. In such a convention, the explicit
form of (A2.1) is given by
$$
(X,Y)\equiv\int\left({\delta_rX\over\delta\phi^i(z)}
\ast{\delta_lY\over\delta\phi_i^\ast(z)}
-{\delta_rX\over\delta\phi_i^\ast(z)}
\ast{\delta_lY\over\delta\phi^i(z)}\right) dz,
\eqno({\rm A}2.2)
$$
where the dualizing density star operator ``$\ast$'' does
not contain the metric tensor, and it takes into account
the total degree rather than the form one. One should
notice that the formula (A2.2) possesses a good geometrical
meaning because the whole integrand is a $d$-form.

The condition of non-degeneracy of a solution of the
master equation assumes, in our convention, the form
$$
\delta_l\delta_r S\vert_{\phi^\ast=0}=\ast Z(x,y),
\eqno({\rm A}2.3)
$$
where $Z(x,y)$ is an integral kernel of gauge
transformation, and the functional differentiation is
performed with respect to an appropriate pair of
(anti)ghosts.$^{10}$
\bigskip
\noindent
{\bf(3)}\quad
It is not absolutely necessary, but technically very
convenient and in accordance with tradition, to introduce
a BRST operator $s$. It follows from the definition of $s$,
$$
(X,S)\equiv sX,
\eqno({\rm A}3.1)
$$
where $S$ satisfies the master equation, that in order to
preserve the self-consistency, we should put
$$
s\phi^I=\ast{\delta_lS\over\delta\phi_I^\ast}
\equiv\ast{\delta S\over\delta\phi_I^\ast},\qquad
s\phi_I^\ast=-\ast{\delta_lS\over\delta\phi^I}
\equiv-\ast{\delta S\over\delta\phi^I}.
\eqno({\rm A}3.2)
$$
Our BRST operator $s$ is nilpotent automatically. This fact
is a consequence of the Jacobi identity
$$
s^2X\equiv((X,S),S)\equiv\half(-)^{\Deg X}((S,S),X)
\equiv\half(-)^{\Deg X}(sS,X)=0,
\eqno({\rm A}3.3)
$$
with the total degree ``Deg'' defined as
$$
\Deg X\equiv\deg X+\gh X,
\eqno({\rm A}3.4)
$$
where ``deg'' is the ordinary form degree, and ``gh'' is
the sum of all four ghost numbers.

There are also some other useful for our further discussion
identities, e.~g.
$$
\{s,d\}\equiv0,
$$
$$
s(XY)\equiv sXY+(-)^{\Deg X}XsY,
$$
$$
XY\equiv(-)^{\Deg X\Deg Y}YX.
\eqno({\rm A}3.5)
$$

\vfill\eject

{\noindent\bf
References}

\frenchspacing
\item{1.} D. Birmingham, M. Blau, M. Rakowski and G.
Thompson, Phys. Rep. 209 (1991) 129.
\item{2.} E. Witten, Commun. Math. Phys. 121 (1989) 351.
\item{3.} G. Horowitz, Commun. Math. Phys. 125 (1989) 417.
\item{4.} G. Horowitz and H. Srednicki, Commun. Math.
Phys. 130 (1990) 83.
\item{5.} I. Oda and S. Yahikozawa, Phys. Lett. B238 (1990)
272.
\item{6.} M. Blau and G. Thompson, Ann. Phys. (NY) 205
(1991) 130.
\item{7.} B. Broda, Phys. Lett. B254 (1991) 111;
\item{} B. Broda, Czech. J. Phys. 42 (1992) 1273.
\item{8.} E. Guadagnini, N. Maggiore and S. P. Sorella,
Phys. Lett. B255 (1991) 65.
\item{9.} M. Blau and G. Thompson, Phys. Lett. B255 (1991)
535;
\item{} B. Broda, Phys. Lett. B280 (1992) 47;
\item{} M. Adud and G. Fiore, Phys. Lett. B293 (1992) 89.
\item{10.} I. A. Batalin and G. A. Vilkovisky, Phys. Rev.
D28 (1983) 2567.
\item{11.} J. C. Wallet, Phys. Lett. B235 (1990) 71;
\item{} H. Ikemori, Mod. Phys. Lett. A 7 (1992) 3397.
\item{12.} L. Baulieu, E. Bergshoeff and E. Sezgin, Nucl.
Phys. B307 (1988) 348;
\item{} M. Abud, J. P. Ader and L. Cappiello, Nuovo Cim.
105A (1992) 1507.
\item{13.} B. Broda, Phys. Lett. B271 (1991) 116.
\item{14.} G. de Rham, Differentiable Manifolds,
Springer-Verlag, Berlin, Heidelberg, 1984, Chapt.~III.

\bye